# A weighted matching scheme of magnetic coil design for FRC shaping control

Zitong Qu[1], Ping Zhu[1,2,*], Zhipeng Chen[1,*], Haolong Li[3], and Jiaxing Liu[1]

[1] State Key Laboratory of Advanced Electromagnetic Technology, International Joint Research Laboratory of Magnetic Confinement Fusion and Plasma Physics, School of Electrical and Electronic Engineering, Huazhong University of Science and Technology, Wuhan, 430074, China

[2] Department of Nuclear Engineering and Engineering Physics, University of Wisconsin-Madison, Madison, Wisconsin 53706, United States of America

[3] College of Physics and Optoelectronic Engineering, Shenzhen University, Shenzhen 518060, China

Email: zhup@hust.edu.cn, zpchen@hust.edu.cn

**Abstract:**

The two-dimentional (2D) separatrix shaping plays a crucial role in the confinement of the Field-Reversed Configuration (FRC), and the magnetic coils serve as an effective means for its control. In this work we develop a method to optimize the location, and current amplitude of the coils required for achieving the target separatrix shape. By iteratively calculating the coil currents, the plasma current, and the equilibrium magnetic flux, the equilibrium separatrix progressively converges towards the desired shape. The coil currents are determined through a matching method, and the NIMEQ code is employed to compute the FRC equilibrium with a Rigid Rotor type of plasma distribution. This approach enables the adaption of the equilibrium separatrix into any desired shape, thus offering a potential coil optimization scheme for the device design and the 2D shaping control of FRC plasma.

# 1. Introduction

The field reversed configuration (FRC) is a promising compact magnetic confinement fusion device that is relatively simple and cost-effective to construct [1,2]. As the FRC vacuum chamber with finite conductivity will eventually lose its ability to conserve magnetic flux, the FRC plasma equilibrium needs to be maintained using magnetic external coils. Meanwhile, the stability of FRC plasma can be significantly influenced by its shaping [3]. Modifying the FRC shaping using external coils is thus a viable method for stability control [4,5], and one of the key priorities to the building of a stable FRC plasma device is the design of the external magnetic coils for the desired equilibrium shaping.

Traditionally, the equilibrium is numerically solved by taking the magnetic flux generated by coil currents at the wall as boundary conditions, utilizing an appropriate plasma pressure as a function of the magnetic flux, and ultimately determining whether the shaping is appropriate [6–8]. The impact of the uncertainty in the boundary conditions on equilibrium necessitates time-consuming adjustments of coil currents and locations, and predicting the effects of changes to the coil configuration on solving equilibrium could be difficult [9]. Xu [10] and Liu [11] adopted a matching approach to construct a Solov'ev tokamak equilibrium with any given 2D shape surrounded by a vacuum region using external poloidal field coils. Their methods require first calculating the analytical equilibrium of the target shape, and then matching to obtain the coil currents and positions. This work develops a weighted matching method for optimising coil design by directly determining the coil currents and locations required for the target separatrix shape from any initial equilibrium, which allows the adaption of the equilibrium separatrix to any desired shape, thus providing a potential coil optimization scheme for device design and the 2D shaping control of FRC plasma.

The rest of this paper is organized as follows. Section 2 introduces the coil optimization method with detailed descriptions of the initialization and the matching procedure. Section 3 presents the results on the application of the method in determining coil currents and locations for several different cases of target separatrix, which is followed by a summary in Section 4.

## 2. Methodology

The cylindrical coordinate system $(R, \Phi, Z)$ is adopted in this paper. The coil design scheme can be outlined using the flowchart in Figure 1. First, for a given target separatrix shape specified as the curve $\vec{R}_{s,tar}(R, Z)$, the initial coil locations and the initial equilibrium flux function $\psi^0$ are assumed. Before each iteration, the weighted average of the separatrix location of the previous iteration $\vec{R}_s^{n-1}$ and the target separatrix location is determined as the matching location $\vec{R}_{ma}^n$. The currents of coils are determined through a matching method outlined in Section 2.2. The equilibrium is then recalculated using NIMEQ by taking the magnetic flux at the edge of the computation domain with contributions from both the coil currents and the plasma current as the boundary condition. The root mean square (RMS) of the $n$-th iterative equilibrium at the target separatrix $\Delta \psi_s = \sqrt{\frac{\sum_{n_s}\left(\psi^n(\vec{R}_{s,tar}) - \psi_s(\vec{R}_{s,tar})\right)^2}{n_s}}$, where $\psi_s(\vec{R}_{s,tar})$ represents the target flux function at $\vec{R}_{s,tar}$, which is conventionally set to be 0, and $n_s$ is the number of selected points on the separatrix. After each iteration, the difference $\Delta\psi_s$ between the iterated flux at $\vec{R}_{s,tar}$ and the target flux $\psi(\vec{R}_{s,tar})$ is evaluated to measure the proximity of the solution to the target value. When $\Delta\psi_s$ decreases below the tolerance $\delta$, the set of coil locations and currents that enables a convergent equilibrium with the target separatrix shape is obtained. More details on the numerical scheme are discussed in the following subsections.

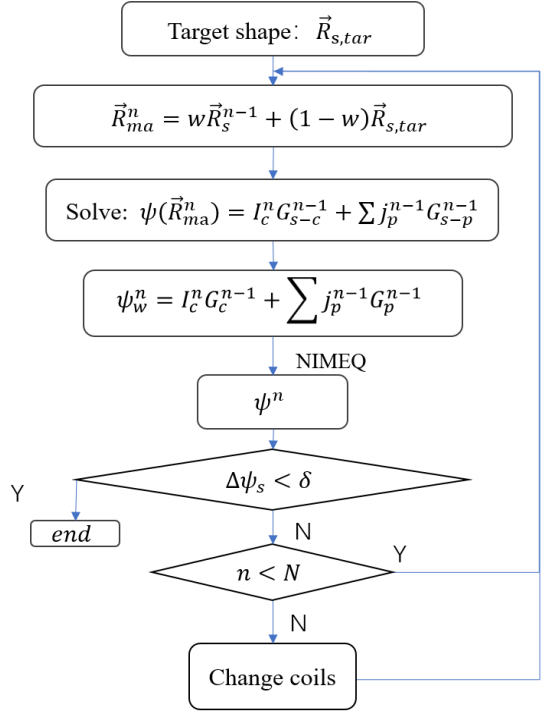

**Figure 1.** Flowchart of the numerical scheme for determining the coil currents. The superscript $n$ in the figure is the number of iterations, $\vec{R}_{ma}^n$ represents the matching location during the $n$-th iteration. $I_c$ and $j_p$ are the coil and plasma current densities respectively, $G$ represents Green function. The value of $\psi(\vec{R}_{ma}^n)$ is assigned to be the same as the flux function on the separatrix, which is set to be zero. $\psi_w$ is the flux function on the domain boundary of NIMEQ determined after matching.

## 2.1 Equilibrium model and solution

The equilibrium solution is obtained by solving the GS equation using the NIMEQ code [12,13]. The Rigid Rotor (RR) model [14,15], which in general agrees better with experiments [16], is chosen as the equilibrium profile for an example. The computation domain of NIMEQ is a rectangular poloidal cross-section with dimensions of 0.35 m × 1.64 m, and the grid is rectangular and uniform. In particular, the GS equation takes the following form

$$\Delta^*\psi = R\frac{\partial}{\partial R}\left(\frac{1}{R}\frac{\partial \psi}{\partial R}\right) + \frac{\partial^2 \psi}{\partial Z^2} = -\mu_0 R j_\Phi \qquad (1)$$

$$j_\phi = e(\omega_i - \omega_e)n_e(0,0)R\, exp\left[\frac{e(\omega_i-\omega_e)}{2T}\psi\right] \qquad (2)$$

where $j_\Phi$ denotes the toroidal current density, $\omega_i$ and $\omega_e$ represent the toroidal rotation frequencies of ions and electrons, respectively. The electron density at the origin of the coordinates is denoted as $n_e(0,0)$, and the electron density is assumed to satisfy the quasi-neutral condition. $\mu_0$ is the vacuum permeability, $e$ represents the elementary charge of electron, and $T$ is the electron temperature, which is assumed to be a constant.

NIMEQ utilizes the weak solution form of the GS equation subject to the boundary condition of the computation domain, which in this study is determined by calculating the boundary magnetic flux $\psi_w$ using the current filament model initially, and the contributions from both the coil and plasma currents later throughout the iteration process.

## 2.2 Method of Green function

The method of matching the magnetic flux at a given separatrix can directly determine the coil current $I_c$. The magnetic flux $\psi_s$ on the separatrix is contributed from the plasma current $j_\Phi$ inside and outside the separatrix, as well as the external coil current $I_c$,

$$\psi_s = \psi_p + \psi_c \tag{3}$$

$$\psi_p(R,Z) = \int_{\Omega_P} dr'dz' G(R,Z,R',Z') j_\Phi(R',Z') \tag{4}$$

$$\psi_c(R,Z) = \sum_{i=1}^{N_c} G(R,Z,R_c,Z_c) I_{c,i} \tag{5}$$

The subscripts $p$ and $c$ represent the plasma and coil, respectively. $G(R,Z,R',Z')$ denotes the Green's function of the plasma at the source point $(R',Z')$ and the field point $(R,Z)$, while $G(R,Z,R_c,Z_c)$ represents the Green's function of the coils at the field point $(R,Z)$. $\Omega_P$ is the plasma poloidal cross-section inside and outside the separatrix. $i$ denotes the serial number of coil groups, with each group consisting of a same number, say ten turns of the same current. Solving equation (3) yields the coil current $I_{c,i}$, which is used next to evaluate the boundary magnetic flux $\psi_w$ at the wall location $(R_w, Z_w)$,

$$\psi_w = \int_{\Omega_P} dr'dz' G(R_w, Z_w, R', Z') j_\Phi(R', Z') + \sum_{i=1}^{N_c} G(R_w, Z_w, R_c, Z_c) I_{c,i} \quad (6)$$

The boundary magnetic flux $\psi_w$ serves as the boundary condition of NIMEQ to solve for the new equilibrium in next iteration.

### 2.3 Weighted matching and iteration

During each iteration, the matching location for determining the coil current is specified as the weighted average of the separatrix location of the previous iterative equilibrium and the location of the target separatrix

$$\vec{R}_{ma}^n = w\vec{R}_s^{n-1} + (1-w)\vec{R}_{s,tar} \quad (7)$$

where $w$ is the weight parameter. The coil current $I_c^n$ is thus obtained by performing magnetic flux matching at the weight-averaged location $\vec{R}_{ma}^n$

$$\psi_s = \psi(\vec{R}_{ma}^n) = I_c^n G_{s-c}^{n-1} + \sum j_p^{n-1} G_{s-p}^{n-1} \quad (8)$$

When RMS value of the iterative equilibrium at the target separatrix, $\Delta\psi_s$ is lower than the predefined tolerance $\delta$, the iteration converges. In this work, the value of $\delta \sim |\psi_{max}^0 \times 10^{-3}|$.

## 3. Applications

### 3.1 Coil configuration optimization

To demonstrate the method, a simple case is selected where the target separatrix is the same as the initial equilibrium with an elongation of 3.23 (Figure 2). The initial coils are symmetrically placed at $R_c = 0.37$ m, $Z_c = (\pm 0.064, 0.191, 0.318, 0.445, 0.572, 0.700)$ m, with each coil having 10 turns. The iteration parameters are set as $\delta = 2 \times 10^{-7}$ Wb, $w = 0.4$, and $N = 10$.

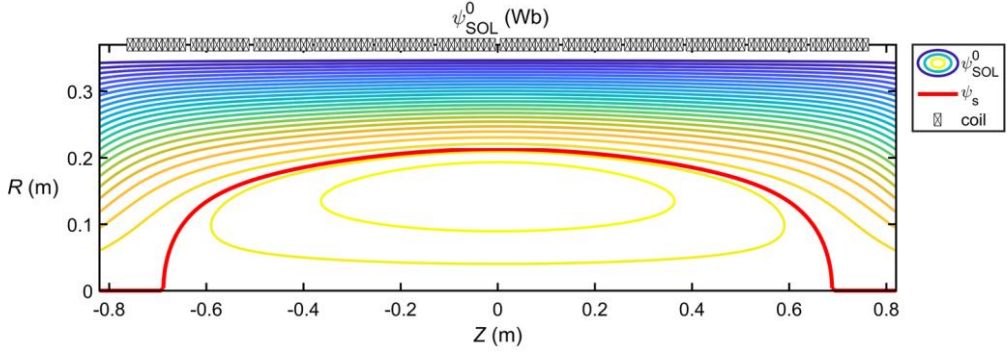

**Figure 2.** Initial equilibrium and coil locations. The separatrix $\vec{R}_s^0 = \vec{R}_{s,tar}$ is denoted as a solid red line, and the coil locations are denoted as squares.

After three iterations, the RMS reaches its minimum but remains higher than $\delta$ (Table 1), and the variations of the separatrix flux at target separatrix $\psi(\vec{R}_{s,tar})$ along the Z-coordinate after each of the three iterations are plotted in Figure 3. The equilibrium with the smallest $\Delta\psi_s$ ($n = 3$) is shown in Figure 4. After iteration, the separatrix of the equilibrium expands outward in the Z-direction with weak magnetic field.

**Table 1.** RMS of per iteration

| $n$ | 1 | 2 | 3 |
|---|---|---|---|
| RMS (Wb) | 6.729e-07 | 5.353e-07 | 4.703e-07 |

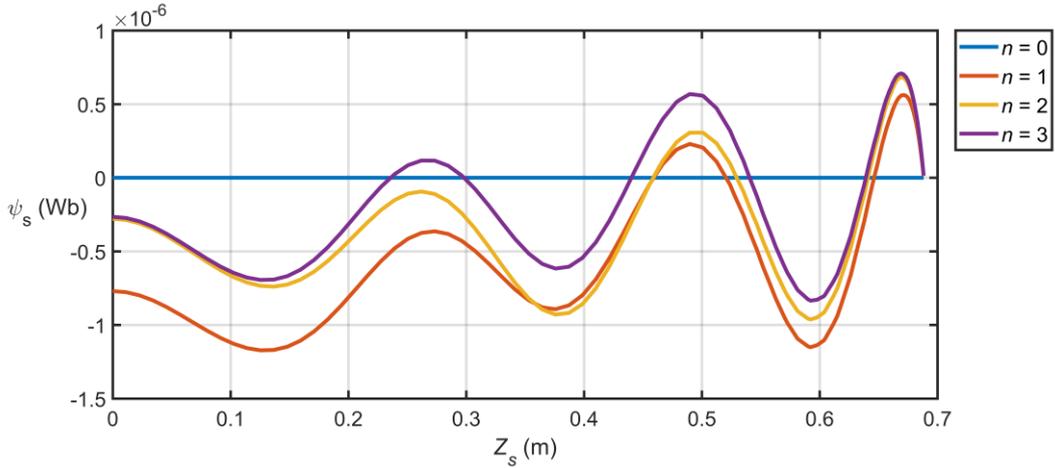

**Figure 3.** Flux function $\psi_s$ at target separatrix along Z-direction after each number of iterations.

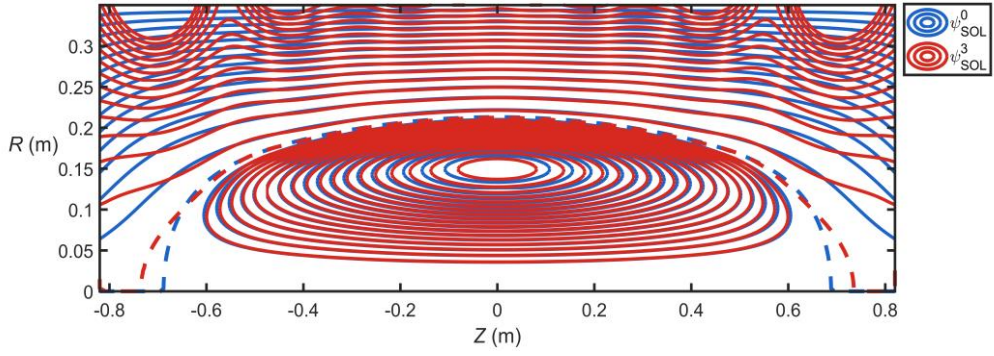

**Figure 4.** Comparison of the $3rd$ equilibrium during iteration and target equilibrium. The dotted lines denote the separatrices, the red solid line denote the final convergent magnetic flux, and the blue solid line denotes the target magnetic flux.

To further reduce the RMS $\Delta\psi_s$, in next step, we add two more coils at $R_c = 0.37$ m and $Z_c = \pm 0.9$ m, and set $w = 0.9$. The RMS values for each iteration (Table 2), the $\psi(\vec{R}_{s,tar})$ along the Z-coordinate during iteration (Figure 5), and comparison between converged and target equilibrium along with the coil locations (Figure 6) indicate convergence.

**Table 2.** RMS of per iteration

| $n$ | 1 | 2 |
| --- | --- | --- |
| RMS (Wb) | 2.227e-07 | 1.246e-07 |

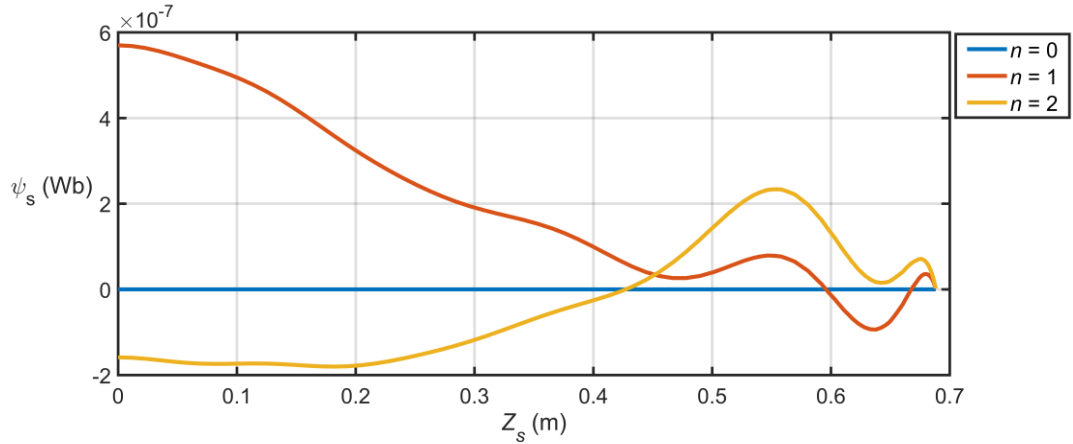

**Figure 5.** Flux function $\psi_s$ at target separatrix along Z-direction after each number of iterations.

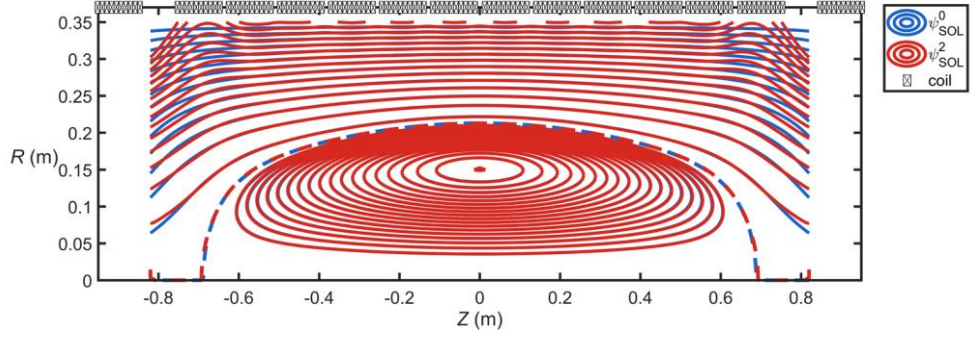

**Figure 6.** Comparison between converged and target equilibrium. The square in the figure shows the location of coils, the dashed line represents the separatrix, while the solid red line depicts the final converged magnetic flux distribution, and the solid blue line denotes the target equilibrium.

However, the current of the 7th ($Z = \pm 0.900$ m) group of coils is 316.67 A per turn, which is about four times higher than other groups, which increases the difficulty of coil manufacturing. Hence, mirror confining coils were used instead, which are placed at $R_c = 0.3$ m, $Z_c = \pm 0.9$ m. The target distribution and location of the equilibrium coil are shown in Figure 7, and the $\psi(\vec{R}_{s,tar})$ at the time of convergence is plotted as the yellow line in Figure 8.

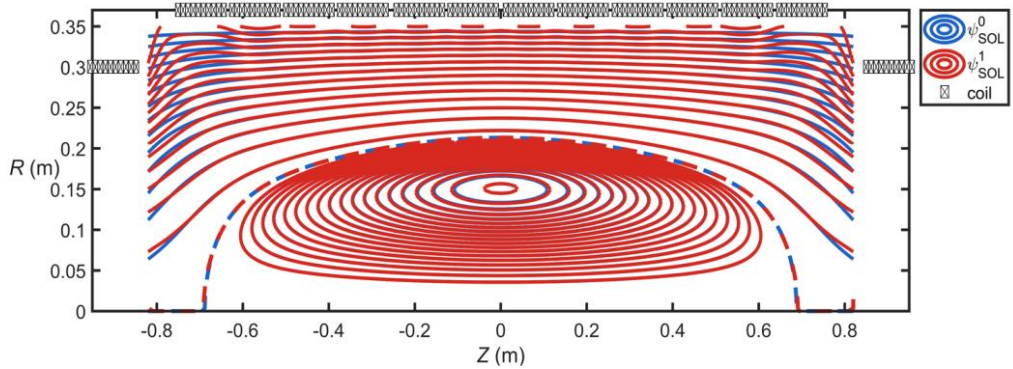

**Figure 7.** Comparison between converged and target equilibrium. The dashed line in the figure represents the separatrix, while the solid red line depicts the final converged magnetic flux distribution, and the solid blue line denotes the target equilibrium.

The iteration converges after one iteration. In Figure 8, the value of $\psi(\vec{R}_{s,tar})$ approaches zero, and the two separatrices in Figure 7 nearly coincide. Furthermore, the current in each turn of the seventh group coil is 237.09 A, which is lower than the previous case and there is no occurrence of current reversal. Notably, there are some difference in the equilibrium distribution near the wall, which can be attributed to the difference between the homogeneous external magnetic field $B_Z$ assumed in the target equilibrium and that from finite number of discrete coils used in calculation.

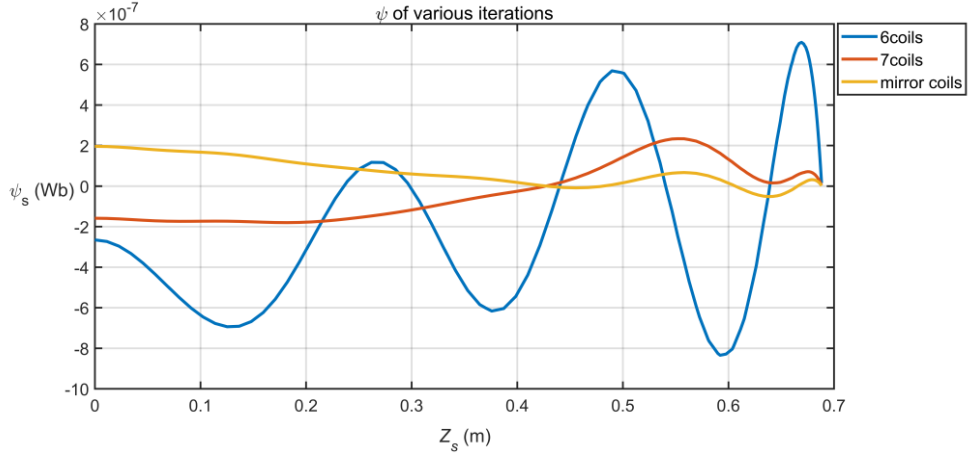

**Figure 8.** Flux function $\psi_s$ at target separatrix along the Z-direction for the three different coil configurations.

Comparison was made among the $\psi(\vec{R}_{s,tar})$ along $Z_s$ for the three different coil distributions, each with the corresponding minimum RMS (Figure 8). Overall, the mirror coil case is the closest to the target separatrix. In addition, for the separatrix near $Z = 0.7$ m, the mirror coil case exhibits the strongest magnetic confinement.

### 3.2 Shaping control

Different from the previous section, where the target separatrix location is assumed same as the initial equilibrium, a target separatrix shape with a smaller elongation of 2.85 is specified to achieve through coil optimization in this section. The mirror coil configuration from the previous section is used for the initial coil locations. After four iterations, the RMS $\Delta\psi_s = 1.577 \times 10^{-7}$ Wb, meets the convergence criterion. The progression of the separatrix location and $\psi(\vec{R}_{s,tar})$ during the iterations presented in Figure 9 (a) and (b), respectively. Throughout the iterations, the shape of the separatrix smoothly transitions from its prolate shape to the desired oblate shape of the target separatrix. The magnetic flux at the target separatrix also converges to zero.

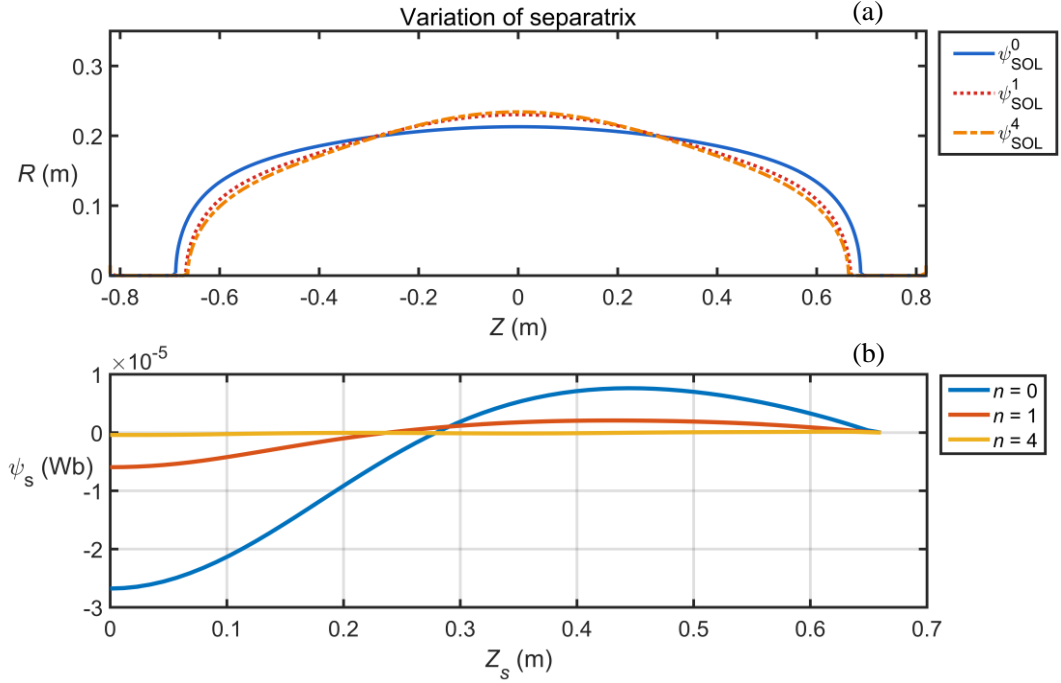

**Figure 9.** (a) Variation of separatrix during the iterations, and (b) the flux function $\psi_s$ at target separatrix along the Z-direction.

The midplane pressure profile of the final converged equilibrium is plotted in Figure 10. The analysis reveals that the overall variation in total pressure ($p + p_b$) is minimal except near the wall, due to the magnetic field curvature.

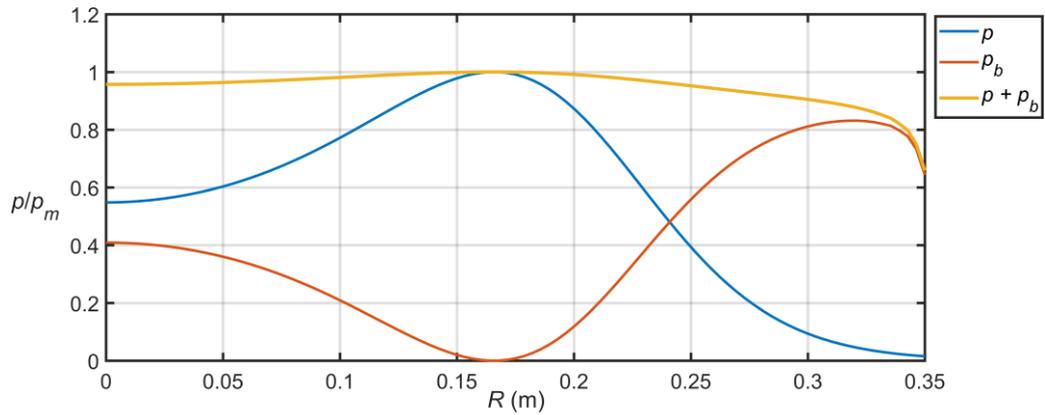

**Figure 10.** Pressure in midplane along R-direction, where $p$ is the plasma pressure, and $p_b$ denotes the magnetic pressure.

Table 3 shows the current values for each group of coils for various elongations. It can be seen from the table that for the equilibrium with smaller elongation, the current values are mostly larger.

**Table 3.** Coil currents with different elongation

| Z location (m) | coil currents of elongation 3.23 (A/turn) | coil currents of elongation 2.85 (A/turn) |
|---|---|---|
| 0.064 | 90.927 | 95.751 |
| 0.191 | 88.399 | 90.124 |
| 0.318 | 87.449 | 87.525 |
| 0.445 | 79.683 | 82.262 |
| 0.572 | 93.387 | 77.305 |
| 0.700 | 48.681 | 82.090 |
| 0.900 | 237.086 | 196.717 |

## 4. Summary

A weighted-matching method has been developed to iteratively determine the coil locations and currents needed to achieve the target separatrix shape. This method enables the determination of coils configuration corresponding to the target separatrix with various elongations. Rapid convergence can be achieved by numerically solving the equilibrium with NIMEQ [12,13], and typically within approximately ten matching iterations. By applying weighted adjustments to the matching positions, the target separatrix is gradually approached, thereby preventing issues arising from discrepancies between the boundary flux of the initial equilibrium and after adjusting the separatrix's shape during numerical equilibrium calculation. The weighted-matching method takes into account the plasma both inside and outside of the separatrix, which is a more realistic representation of FRC plasma in comparison to other models [10,11,17].

Nevertheless, in this study the equilibrium outside the separatrix still suffers the unrealistic and excessive width of the scraping off layer from the RR model. In addition, it is uncertain if this method remains effective or applicable in presence of a relatively large equilibrium flow. We intend to examine these issues and explore the potential of the method as a real-time shaping control scheme in future work.

## Acknowledgments

The authors are very grateful for the help of the HFRC team in the State Key Laboratory of Advanced Electromagnetic Engineering and Technology of China and the support from the NIMROD team. This work was supported by the National Key Research and Development Program of China (Grant No. 2017YFE0301804), the Fundamental Research Funds for the Central Universities at Huazhong University of Science and Technology (Grant No. 2019kfyXJJS193), the National Natural Science Foundation of China (Grant No. 51821005), and the U.S. Department of Energy (Grant Nos. DE-FG0286ER53218 and DE-SC0018001). The computing work in this paper was supported by the Public Service Platform of High Performance Computing by Network and Computing Center of HUST.

# Reference

[1] Tuszewski M 1988 Field reversed configurations *Nucl. Fusion* **28** 2033–92

[2] Steinhauer L C 2011 Review of field-reversed configurations *Phys. Plasma* **18** 070501

[3] Cobb J W, Tajima T and Barnes D C 1993 Profile stabilization of tilt mode in a field-reversed configuration *Phys. Fluids B* **5** 3227–38

[4] Gota H, Binderbauer M W, Tajima T, Smirnov A, Putvinski S, Tuszewski M, Dettrick S, Gupta D, Korepanov S, Magee R, Park J, Roche T, Romero J, Trask E, Yang X, Yushmanov P, Zhai K, Schmitz L, Lin Z, Ivanov A, Asai T, Baltz T and Platt J 2021 Overview of C-2W: High-temperature, steady-state beam-driven field-reversed configuration plasmas *Nucl. Fusion* **61** 106039

[5] Myers C E, Edwards M R, Berlinger B, Brooks A and Cohen S A 2012 Passive Superconducting Flux Conservers for Rotating-Magnetic-Field-Driven Field-Reversed Configurations *Fusion Sci. Technol.* **61** 86–103

[6] Ma H J, Xie H S, Bai Y K, Cheng S K, Deng B H, Tuszewski M, Li Y, Zhao H Y, Chen B and Liu J Y 2021 Two-parameter modified rigid rotor radial equilibrium model for field-reversed configurations *Nucl. Fusion* **61** 036046

[7] Takahashi T, Gota H and Nogi Y 2004 Control of elongation for field-reversed configuration plasmas using axial field index of a mirror confinement field *Phys. Plasma* **11** 4462–7

[8] Lackner K 1976 Computation of ideal MHD equilibria *Comput. Phys. Comm.* **12** 33–44

[9] Guazzotto L 2017 Coil current and vacuum magnetic flux calculation for axisymmetric equilibria *Plasma Phys. Control. Fusion* **59** 122001

[10] Xu T and Fitzpatrick R 2019 Vacuum solution for Solov'ev's equilibrium configuration in tokamaks *Nucl. Fusion* **59** 064002

[11] Liu J, Zhu P and Li H 2022 Two-dimensional shaping of Solov'ev equilibrium with vacuum using external coils *Phys. Plasma* **29** 084502

[12] Howell E C and Sovinec C R 2014 Solving the Grad–Shafranov equation with spectral elements *Comput. Phys. Comm.* **185** 1415

[13] Li H and Zhu P 2021 Solving the Grad–Shafranov equation using spectral elements for tokamak equilibrium with toroidal rotation *Comput. Phys. Commun.* **260** 107264

[14] Morse R L 1970 Rigid Drift Model of High-Temperature Plasma Containment *Phys. Fluids* **13** 531

[15] Qerushi A and Rostoker N 2002 Equilibrium of field reversed configurations with rotation. III. Two space dimensions and one type of ion *Phys. Plasma* **9** 5001–17

[16] Armstrong W T 1981 Field-reversed experiments (FRX) on compact toroids *Phys. Fluids* **24** 2068

[17] Guazzotto L 2017 Coil current and vacuum magnetic flux calculation for axisymmetric equilibria *Plasma Phys. Contr. F.* **59** 122001